\documentclass[a4paper,final]{article}
\usepackage{latexsym,amsmath,amssymb,showkeys,psfig}

\newcommand{\scri}{\mathcal{J}}

\newcommand{\cE}{\mathcal{E}}

\newcommand{\del}{\partial}

\newtheorem{definition}{Definition}

\numberwithin{equation}{section}

\title{%
Numerical treatment of the hyperboloidal initial value problem for the
vacuum Einstein equations. \\
III. On the determination of radiation
}
\author{J\"org Frauendiener\\
Institut f\"ur Theoretische Astrophysik,\\
Universit\"at T\"ubingen,\\
Auf der Morgenstelle 10,\\
D-72076 T\"ubingen,\\
Germany}

\begin{document}
\maketitle

\begin{abstract}
We discuss the issue of radiation extraction in asymptotically flat
space-times within the framework of conformal methods for numerical
relativity. Our aim is to show that there exists a well defined
and accurate extraction procedure which mimics the physical
measurement process. It operates entirely intrisically within
$\scri^+$ so that there is no further approximation necessary apart
from the basic assumption that the arena be an asymptotically flat
space-time. We define the notion of a detector at infinity by
idealising local observers in Minkowski space. A detailed discussion is 
presented for Maxwell fields and the generalisation to linearised and
full gravity is performed by way of the similar structure of the
asymptotic fields. 
\end{abstract}

\section{Introduction}
\label{sec:intro}

In~\cite{jf-1997-2} and~\cite{jf-1997-3} we have described a numerical
approach towards the solution of the Einstein vacuum equations for
asymptotically flat spacetimes. It is based on Friedrich's conformal
field equations~\cite{Friedrich-1981} which in turn are inspired from
Penrose's geometric ideas. In particular his conformal
compactification procedure of asymptotically flat space-times allows
to bring infinitely distant points in to finite regions. The
advantages and disadvantages of this ``conformal method'' have been
discussed at various
places~\cite{jf-1997-2,jf-1997-3,Huebner-1998-2}. Here we want to
focus on one particular issue, namely how to extract radiation
information from the numerically generated data and how to interpret
them. This is not intended to be a detailed list of the routine steps
to get to the relevant information because this has already been given
in~\cite{jf-1997-3}. Rather, we want to discuss this problem as a
matter of principle, the emphasis of the discussion being on the
mathematical idealizations which go into the process. 

Our motivation for doing so is mainly to clarify. People have raised
some doubts in the past whether radiation extraction in the context of
conformal space-times can be a faithful process because the conformal
compactification distorts time and space in a rather violent way thus
giving rise to inaccurate frequencies and wavelengths of the radiation
registered by a detector. Our aim in this paper is to present
convincing arguments that, in fact, the conformal compactification
works so beautifully that the extraction procedure based on the
conformal method is by way of principle more faithful than procedures
in use in standard numerical approaches.

Certainly, most of the ideas presented in this work are not new. They
are probably all hidden in the early works on the asymptotic structure
of the zero rest mass fields. However, it is important to collect them
here in one place because the issues which arise in this new conformal
approach are to a large extent unknown to the numerical relativity
community.

The plan of the paper is as follows: we start out with a very simple
radiating system, the Hertz dipole. We isolate the interesting
quantity which is to be detected and show which field component is
relevant. In the next section we discuss detectors and in particular
how to idealise a detector at infinity. This notion is then used to
demonstrate how to extract the dipole radiation on Minkowski space
using conformal methods. Finally, we show how these ideas can be
applied to linearized gravity and to the full non-linear theory.

\section{Radiation in Maxwell theory}
\label{sec:radmaxwell}

``Radiation'' is a rather obscure notion in physics. Before discussing
things in gravitational theory let us look at a simpler theory,
Maxwell theory. Even there it is not easy to give a straightforward
definition of that term. The problem is that ``radiation'' is a global
concept but it is often referred to in a local context. One possible
approach towards electromagnetic radiation is to consider a system as
radiating if there is a nonvanishing energy flux through the sphere at
infinity. In this statement, there is a clear reference to
infinity. Another instance of the necessity of infinity for radiation
is the fact that the radiation fields which are obtained from time
varying multipoles are defined in the ``wave zone'' which in turn is
defined as the region where the distance from the source is much
larger than the characteristic size of the source. Hence the radiaton
fields are obtained as a limit $r\to\infty$ of the full field. In
order to have an unambiguous definition of radiation one has to have
access to the limit $r\to \infty$. It is very important to state
exactly how this limit is to be taken. Without specifying what happens 
to the other, especially the time coordinate, this limit may lead to
unreasonable results. Thus, one has to study radiation from a global
space-time point of of view. 

To get a feeling for what happens let us start out with a very simple
example of a radiating system in Maxwell theory, the Hertz dipole. As
our starting point, we take the relevant formulae for the fields of a
radiating dipole from Jackson~\cite{Jackson-1975}:
\begin{align}
  \mathbf{\hat B}(\omega,\mathbf{r})&= k^2 (\mathbf{n\times
    \hat{p}}(\omega) \frac{e^{ikr}}{r}  
  \left(1-\frac1{ikr}\right),\\
  \mathbf{\hat E}(\omega,\mathbf{r})&= k^2 \mathbf{(n\times
    \hat{p}(\omega))\times \hat{p}(\omega)} 
  \frac{e^{ikr}}{r} \\
  &+ \left[3\mathbf{n(n\cdot \hat{p}(\omega))-\hat{p}(\omega)}\right]
  \left(\frac1{r^3} 
    -\frac{ik}{r^2} \right)e^{ikr}.\nonumber
\end{align}
Here, $\mathbf{n} = \mathbf{r}/r$ and all hatted quantities are
supposed to have a periodic time dependence $e^{-i\omega t}$, so they
are functions of the frequency $\omega$. In addition, the fields
$\mathbf{\hat E}$ and $\mathbf{\hat B}$ depend on the position vector
$\mathbf{r}$. Unfortunately, the representation of the fields in
terms of their Fourier transforms in time is totally inadequate for
space-time considerations. Therefore, we undo the Fourier decomposition
and write the fields in Minkowski coordinates $(t,\mathbf{r})$. Taking
into account the dispersion relation $\omega^2 - k^2=0$ and choosing
$k=\omega$ for retarded fields we obtain for
the magnetic field
\[
  \mathbf{B}(t,\mathbf{r}) = \int \mathbf{\hat B}(\omega,\mathbf{r})\,
  e^{-i\omega t}\,d\omega = \int \left(\frac{\omega^2}r +
  \frac{i\omega}{r^2} \right) \mathbf{(n \times \hat p(\omega))}\,
  e^{-i\omega (t-r)}\,d\omega .
\]
Then, this expression and a similar one for the electric field result
in the following electromagnetic dipole fields
\begin{align}
  \mathbf{B}(t,\mathbf{r}) &= - \frac1r (\mathbf{n\times\ddot p}(t-r)) -
  \frac1{r^2}(\mathbf{n\times\dot p}(t-r)),\\
  \mathbf{E}(t,\mathbf{r}) &= - \frac1r (\mathbf{n\times \ddot p}(t-r))
  \times \mathbf{n} + 
  \frac1{r^3} \left[3\mathbf{n}(\mathbf{n\cdot {p}}(t-r)) -
    \mathbf{p}(t-r)\right] \nonumber \\
  & + 
  \frac1{r^2} \left[3\mathbf{n}(\mathbf{n\cdot \dot{p}}(t-r)) -
    \mathbf{\dot{p}}(t-r)\right] . 
\end{align}
We use the convention that here and in the sequel a dot will always
mean the derivative of a function of one variable with respect to its
argument. So here, $\mathbf{p}$ depends on $u=t-r$ and the dot stands for
$d/du$. It represents a dipole moment which is located at $r=0$ and
which varies in time. This form of the fields could have been obtained 
by applying the Li\'enard-Wiechert potentials to the case of a dipole
and computing the fields from them (see e.g.,
Sommerfeld~\cite{Sommerfeld-1977}).

In order to simplify the formulae slightly, without loosing much
generality we require $\mathbf{p}$ to be of the form
$p(u)\mathbf{e}_z$, i.e., to be always pointing in a fixed direction,
the $z$-axis, thus producing an axisymmetric field
configuration. Introducing polar coordinates $(r,\theta,\phi)$ and the
three unit vectors $\mathbf{e}_r=\del_r=\mathbf{n}$,
$\mathbf{e}_\theta = 1/r\cdot \del_\theta$ and $\mathbf{e}_\phi =
1/(r\sin\theta)\cdot \del_\phi$ we get the simpler formulae
\begin{align}
  \mathbf{B}(t,\mathbf{r}) &= \sin\theta \left(\frac{\ddot p(u)}{r} +
    \frac{\dot p(u)}{r^2} 
  \right)\, \mathbf{e}_\phi,\\ 
  \mathbf{E}(t,\mathbf{r}) &= \sin\theta
  \left(\frac{\ddot p(u)}{r} + \frac{\dot p(u)}{r^2} +
    \frac{p(u)}{r^3}\right)\,\mathbf{e}_\theta 
  + 2 \cos\theta \left(\frac{\dot p(u)}{r^2} + \frac{p(u)}{r^3} \right)
  \,\mathbf{e}_r.
\end{align}
This is the general field configuration for a dipole which is oriented
along a fixed axis. It includes both the radiative fields and the
field in the ``near zone''. 

To extract the radiative part we consider the Poynting vector
$\mathbf{S=E\times B}$. Its flux integral over a closed 2-dimensional
surface represents the energy which flows through that surface.  We
obtain
\[
\mathbf{S}(t,\mathbf{r}) = \sin^2\theta \left(\frac{\ddot p(u)}{r}\right)^2
\mathbf{e}_r + O(r^{-3}) 
\]
The exhibited term is the only one which survives integration over a
sphere with radius $R$ in the limit $R\to \infty$. Thus, the radiated
power of the system is
\[
\lim_{R\to\infty}  \int \mathbf{(S\cdot n)}\,R^2\sin\theta d\theta d\phi 
= \frac{8\pi}3 \lim_{R\to\infty}{\ddot p(t-R)}^2 
\]
Already, we find that one has to specify what to do with the
$t$-coordinate when taking the limit. Simply taking $t$ to be constant 
results in an uninteresting limit with $\ddot p$ being evaluated at
$u=-\infty$. We will discuss this issue in detail below.

The radiation field consists of the terms which are proportional to
$\ddot p$ and these are the ones which fall off as $1/r$:
\begin{align}
  \mathbf{B}_{\mbox{rad}}(t,\mathbf{r}) &= - \frac1r
  (\mathbf{n\times\ddot p}(u)) = 
  \sin\theta \,\frac{\ddot p(u)}{r}\, \mathbf{e}_\phi,\\  
  \mathbf{E}_{\mbox{rad}}(t,\mathbf{r}) &= - \frac1r
  (\mathbf{n\times\ddot p}(u))  \times
    \mathbf{n} = \sin\theta\, 
  \frac{\ddot p(u)}{r}\,\mathbf{e}_\theta.
\end{align}
These are also the ones which one would obtain by solving the
Maxwell equations for large $r$. By themselves they are a solution of
the Maxwell equations.

Now, what is the ``radiative information'' we are
interested in? Obviously, the only
information contained in the radiative fields is the function $\ddot
p(u)$. Let us agree, that it is this function which we are interested
in, because by analysing the far-field, extracting $\ddot p$, we can
obtain information about the structure of the source. In fact, we
will show below that this is (almost) the so called ``news
function''. 

It should be emphasized that the function $\ddot p$ here plays two
roles. On the one hand, it describes the behaviour of the source,
namely the time variation of the dipole, and on the other hand, it
characterizes the asymptotic fields in the form of the news
function. In the present case of Maxwell theory, both these properties
coincide and there is a simple unique relationship between the
far-field and the source. While this is still the case in linearized
gravity it is no longer true in the full non-linear theory of
gravity. It is expected that there will still be a unique relationship
between the radiation fields and the structure (multipole moments) of
the source but certainly it will not be as direct as in our simple
example because during the way from the source to infinity the fields
will interact with themselves due to the non-linear nature of the
theory.

\section{The detector at infinity}
\label{sec:detect}

The next question to consider is how to extract the information. In
real life this is achieved by an antenna or a detector which uses the
interaction with the fields to get information about their strength in
various directions of space. Let us idealize such an antenna by a
freely falling (non-rotating) reference frame, i.e., a timelike
geodesic parametrized by proper time which has attached to it a triad
of spacelike unit-vectors orthogonal to the timelike tangent and
transported along the geodesic according to Fermi-Walker. The
measuring process will be idealized as a multiplication of the fields
with $r$ followed by a simple projection onto the three axes. Then we
take a detector far away from the source , i.e., $(r,\theta,\phi)$
fixed with $r$ large, and try to determine the information. Since we cannot
separate the far-field from the near-field we have to measure the full
field at the location of the antenna and we find that we cannot
extract a clean news function because there will be admixtures of the
$1/r^2$ and higher terms to the measurement result. Therefore, we need
to take the antenna very far away from the source and, ideally, we
need to take the antenna out to infinity.

To see more clearly what happens we now view this situation in a
space-time which is conformal to Minkowski space, the Einstein
cylinder $\cE$. To perform the conformal transformation we introduce
null coordinates $u=t-r$ and $v=t+r$. This puts the Minkowski
line-element into the form
\begin{equation}
  \tilde g = du\,dv - \frac14(v-u)^2d\sigma^2,
\end{equation}
where $d\sigma^2$ is the metric of the unit-sphere. The coordinates
$u$ and $v$ each range over the complete real line, subject only to
the condition $v-u\ge0$. This infinite range is compactified by
transforming with an appropriate function, e.g.,
\begin{equation}
  u = \tan U,\qquad v=\tan V,
\end{equation}
thus introducing new null coordinates $U$ and $V$, in terms of which
the metric takes the form
\begin{equation}
  \tilde g = \frac{1}{4\cos^2U\cos^2V} \left\{ 4dU\,dV -
    \sin^2(V-U)\,d\sigma^2 \right\}. 
\end{equation}
The coordinates $U$, $V$ both range over the open interval
$({-\frac\pi2},\frac\pi2)$ with the restriction $V-U\ge0$. Obviously,
the Minkowski metric becomes singular at points with $U=\pm\frac\pi2$
or $V=\pm\frac\pi2$.

Now we define a different metric $g=\Omega^2\tilde g$, conformally
related to $\tilde g$ by the conformal factor $\Omega=2\cos U\cos
V$. Thus,
\begin{equation}
   g =  4dU\,dV - \sin^2(V-U)\,d\sigma^2.
\end{equation}
This metric is perfectly regular at the points mentioned above and, in
fact, $g$ is the metric of the Einstein cylinder. This can be verified
easily by defining an appropriate time and radius coordinate,
see~\cite{HawkingEllis-1973}.

\begin{figure}[htbp]
  \begin{center}
    \begin{picture}(60,60)
      \put(25,5){\mbox{\psfig{file=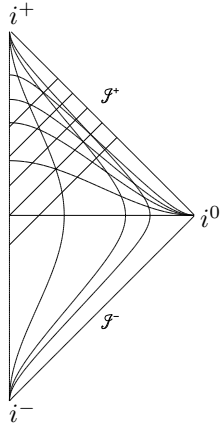,height=5cm}}}
      \put(52,30){\makebox(0,0){$i^0$}}
      \put(27,57){\makebox(0,0){$i^+$}}
      \put(27,4){\makebox(0,0){$i^-$}} 
    \end{picture}
    \caption{The  conformal diagram of Minkowski space. The
      lines meeting at $i^0$ are lines of constant $t$, while the
      lines emanating from $i^-$ and converging into $i^+$ are lines
      of constant $r$. They are the world-lines of detectors at a
      finite distance from the source. The lines at $45^\circ$
      symbolise radiation.}
    \label{fig:mink}    
  \end{center}
\end{figure}

\begin{figure}[htbp]
  \begin{center}
      \mbox{\psfig{file=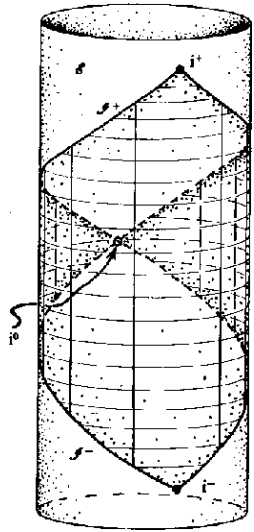,height=7cm}}
    \caption{The embedding of Minkowski space into the Einstein
      cylinder $\cE$. (This figure is taken
      from~\cite{PenroseRindlerII})} 
    \label{fig:emb}
  \end{center}
\end{figure}

In Fig.~\ref{fig:mink} is shown the standard conformal diagram
for Minkowski space~\cite{Penrose-1965}.
Each point in the interior of the triangle corresponds to a
2-sphere. The long side of the triangle consists of all the points in
the center, $r=0$ (i.e., $U=V$). The other two sides of the triangle
correspond to the points with $V=\frac\pi2, |U|<\frac \pi2$
($\scri^+$) and $U=-\frac\pi2, |V|<\frac \pi2$ ($\scri^-$). These are
3-dimensional null-hypersurfaces which represent (future and past)
``null-infinity''. The points $i^\pm$ are points in the center with
$U=V=\pm \frac \pi2$, while $i^0$ is a point with $U=-\frac \pi2,
V=\frac \pi2$. These three points represent future and past timelike
infinity and spacelike infinity. Thus, we may consider Minkowski space
to be conformally embedded into the Einstein cylinder. This is shown in
Fig.~\ref{fig:emb}. For more information on general asymptotically
flat space-times and their conformal extensions
see~\cite{PenroseRindlerII}.

The characteristic feature of the dipole field is, of course, that it
is a retarded field: if $p$ is such that $\ddot p(u)=0$ for $u\le0$
then no detector will see anything at times $t$ with $u=t-r\le0$. The
information travels along the hypersurfaces $u=\mathrm{const.}$, the
future light cones emanating from the world-line of the source. These
are shown in fig.~\ref{fig:mink} as straight lines reaching out to
$\scri^+$.

What happens now if we take a detector and move it further and further 
out to infinity in order to extract the news function as cleanly as
possible? This is not as straightforward as one would expect. The
world-line of a detector is given by (ignoring the angular variables)
\[
\gamma_P(\lambda)=(t(\lambda),r(\lambda)) = (T+\lambda,R).
\]
The event $P=\gamma_P(0)$ which characterizes the world-line of the
detector uniquely can be interpreted as the event when the detector
``switches on'', i.e., starts to register.  We have $T=t(P)$ and
$R=r(P)$. The world-line of the detector in terms of
$(u,v)$-coordinates is
\begin{equation}
  \gamma_{P}: (u(\lambda),v(\lambda)) = (T-R +\lambda,T+R
  +\lambda) 
\end{equation}
and in terms of $(U,V)$-coordinates one has
\begin{equation}
  \gamma_{P}: (U(\lambda),V(\lambda)) = (\arctan(T-R +\lambda),
  \arctan(T+R + \lambda) ) 
\end{equation}
We take a sequence of detectors, determined by their ``switch-on
events'' $P_n$ (with $T_n=t(P_n)$ and $R_n=r(P_n)$) such that $R_n \to
\infty$ for $n\to \infty$. First suppose the sequence is such that
$T_n=T=\mathrm{const}$. Then we obtain a ``limiting detector''
\begin{equation}
  \lim_{n\to\infty} \gamma_{P_n}(\lambda)= \lim_{n\to\infty}
  (U_n(\lambda),V_n(\lambda)) = (-\frac\pi2,\frac\pi2),
\end{equation}
i.e., the limiting detector has been compressed entirely to the point
$i^0$. This limit is reasonable but not useful. In fact, what happens
is that the detectors switch on at the same time $t=T$. Then they have
to sit and wait for the signal to arrive, the delay growing longer and
longer, depending on their distance to the source, and in the limit
there will be no signal at all.

Therefore, we need to use a different sequence of detectors.
Operationally speaking, we trigger the detectors to switch on when the
signal arrives. Thus, we have a sequence for which
$T_n-R_n=u_0=\mathrm{const}$. Then we obtain
\begin{equation}
  \lim_{n\to\infty} \gamma_{P_n}(\lambda)= \lim_{n\to\infty}
  (U_n(\lambda),V_n(\lambda)) = (\arctan(u_0+\lambda),\frac\pi2).
\end{equation}
This ``limiting detector'' is a line on $\scri^+$ determined by the
angles $\theta$ and $\phi$. Furthermore, it is a generator of
$\scri^+$ and, hence, a null geodesic. However, the parameter
$\lambda$ is not just any parameter along that geodesic. Instead,
it is a so called Bondi parameter~\cite{PenroseRindlerII}: let $s$ be
an arbitrary parameter along the generator and let $\mathbf{k}=d/ds$
be the tangent vector along the generator with $\mathbf{k}(s)=1$. The
acceleration $a$ of $\mathbf{k}$ is defined by the equation
$d\mathbf{k}/ds=a\mathbf{k}$. Then $\lambda$ is a Bondi parameter if
it satisfies the equation
\begin{equation}
\label{Bondiparm}
  \left(\frac{d}{ds} - a - 2\rho'\right)\frac{d\lambda}{ds} =0.
\end{equation}
Here, $\rho'$ is the expansion of $\scri^+$, a quantity which
determines how quickly neighbouring generators spread apart. It is
determined intrinsically from the way $\scri^+$ is attached to the
physical space-time. The concept of a Bondi parameter is conformally
invariant so it does not depend on which conformal factor has been
used to attach the boundary to physical space. If $\lambda$ is a Bondi
parameter then every other Bondi parameter has the form
$G(\theta,\phi) \lambda + H(\theta,\phi)$ where $G$ and $H$ are
arbitrary functions on the sphere of generators of $\scri^+$. 

This freedom in the choice of parameters is exactly what one would
expect considering the way it was defined. Recall that we took the
sequence of detectors in such a way that their world-lines were
defined by keeping the spatial coordinartes fixed. However, these
coordinates were one special choice of Minkowski coordinates. Any
other choice would serve the same purpose. Thus, we have the freedom
of the full Poincar\'e group available to define the detectors. The
limit of these detectors will always be a generator of $\scri^+$ with
a Bondi parameter on it but the parameters will not be the
same. Instead they will exhaust the full 2-dimensional affine space of
possible Bondi parameters.

To verify that $\lambda$ is indeed a Bondi parameter in our case we
observe that $U$ can be taken as a parameter along the generator. In
fact, it is an affine parameter so that the corresponding acceleration
vanishes. For the expansion we get $\rho'=\tan U$. Inserting this into
the equation \eqref{Bondiparm} shows that it is identically
satisfied. In fact, we may take the retarded time $u=\tan U$ itself as
a Bondi parameter because it also satisfies \eqref{Bondiparm}. In
order to complete the notion of a detector at infinity we observe that
the radiation fields are both tangent to the spheres of constant $t$
and $r$ and this remains so even in the limit $r\to \infty$. Thus, we
require that the detector at infinity has two of its axes tangent to
$\scri^+$ and perpendicular to the generator.

What we have found with this limiting procedure is a concept which is
entirely intrinsic to $\scri^+$. It does not depend anymore on the
fact that we have started with Minkowski space. Instead it is a
concept which can be introduced whenever the notion of null-infinity
is well defined. Therefore, we now make the 
\begin{definition}
  A detector at infinity is idealized by a generator of $\scri^+$
  which is parametrized by a Bondi parameter $u$. The axes of the detector 
  should be oriented so that one axis (the ``retarded time'' axis)
  coincides with $d/du$ while two other axis are perpendicular to the
  generator and tangent to $\scri^+$.
\end{definition}

\section{Radiation extraction at infinity}
\label{sec:radinf}

Let us see whether we can now use this definition to extract radiative
information. Suppose we were to solve the Maxwell equations
numerically in Minkowski space. And suppose we do not use the
traditional way but instead solve a conformally equivalent set of
equations on a conformally related space like the Einstein cylinder
$\cE$. The Maxwell equations are conformally invariant so we still
have to solve the Maxwell equations, only reexpressed in terms of the
conformal metric $g$ instead of the Minkowski metric $\tilde g$. Of
course, we have to put in somewhere that we are really interested in
fields on Minkowski space and not in the more general fields on the
Einstein cylinder $\cE$. This is done by specifying the initial data
appropriately.  

The numerical solution of these equations provides us with a Maxwell
field on the complete Einstein cylinder including that piece which
corresponds to Minkowski space and its conformal boundary
$\scri^+$. Usually, this solution will be represented on 3-dimensional
spacelike hypersurfaces which form a foliation of the conformal
space. On each such slice, we can now search for the location of its
intersection with $\scri^+$, a 2-dimensional closed surface. Then we
can evaluate the Maxwell field on that surface and so, keeping the
angles fixed and going from one slice to the next, we obtain the value
of the Maxwell field along a generator of $\scri^+$ as a function of
the time parameter which labels the slices.

As an example, we consider again the dipole fields. In order to get
the solutions onto the Einstein cylinder we first write the electric
and magnetic field as one-forms and then combine them into the Faraday 
2-form according to
\begin{equation}
  F=dt\wedge E + \star B.
\end{equation}
The star is the Hodge star of the metric on the euclidean spaces
$t=\mathrm{const}$. The covariant form of the fields turns out to be
\begin{align}
  E &= \left(\ddot p + \frac{\dot p}{r} + \frac{p}{r^2}\right)\,
  \sin\theta\, d\theta + 2 \cos\theta\left(\frac{p}{r^3} + \frac{\dot
      p}{r^2}\right)\, dr,\\
  B &= \left( \ddot p + \frac{\dot p}{r} \right)\, \sin^2\theta\,d\phi.
\end{align}
In terms of the double null coordinates $u$ and
$v$ we obtain 
\begin{equation}
  F= -\left( \ddot p + \frac{\dot p}{r} \right)\, du\wedge dy
  - \frac{p}{2r^2}\left(du + dv\right)\wedge dy 
  + y\left(\frac{p}{r^3} + \frac{\dot p}{r^2}\right)\,du \wedge dv,
\end{equation}
where $r=(v-u)/2$ is considered as a function of $u$ and $v$ and where
we have introduced $y=\cos\theta$ as another coordinate. Observe, that
the radiation field sits entirely in the $du\wedge dy$ component of
$F$. It is the only component which survives the limit $r\to \infty$,
keeping $u$ fixed. In the non-axisymmetric case we would also get a
radiation component proportional to $du\wedge d\phi$ and this is the
general behaviour for Maxwell fields on asymptotically flat
space-times. This implies that we get the radiation fields from the
Faraday 2-form simply by restricting it to $\scri^+$. 

The Faraday 2-form is conformally invariant. This implies that one
gets the corresponding Faraday form on the Einstein cylinder by
performing the coordinate transformation which transforms from the
$u,v$ coordinates to the $U,V$ coordinates. This results in
\begin{align*}
  F&= -\left( \left(\ddot P -2 \tan U \dot P\right) \cos^2 U + 2
    \frac{\dot P \cos U \cos V}{\sin(V-U)} \right)\, dU\wedge dy \\
  &- \frac{P}{2\sin^2(V-U)}\left(\cos^2V dU +  \cos^2U dV\right)\wedge dy \\
  &+ 4y\left(\frac{2P\cos U \cos V}{\sin^3(V-U)} + \frac{\dot P
      \cos^2U}{\sin^2(V-U)}\right)\,dU \wedge dV, 
\end{align*}
where the function $P$ is defined by $P(U)=p(\tan U)$ and the dot here 
means $d/dU$.

Let us now restrict this form to $\scri^+$ by pulling it back along
the inclusion map 
$i:\scri^+ \to \cE, (U,y,\phi)\mapsto (U,\frac\pi2,y,\phi)$ 
\begin{equation}
  i^*F=-\left(\ddot P -2 \tan U \dot P\right) \cos^2 U \, dU\wedge dy .
\end{equation}
According to the discussion above, this is the radiation field.

The detector at infinity is a line $y=\mathrm{const}.$ (and
$\phi=\mathrm{const.}$) on $\scri^+$ with a Bondi parameter on it. The
variable $U$ is not a Bondi parameter. In order to find one we need to
solve the second order equation~\eqref{Bondiparm}. Of course, we know
that $u=\tan U$ is a solution, so that the radiation field in terms of
this parameter reads
\begin{equation}
  i^*F=- \ddot {\tilde p}(u) \, du\wedge dy ,
\end{equation}
where $P(U)=\tilde p(\tan U)$. By comparison with the above definition 
of $P$ we get $p(u)=\tilde p(u)$. Hence, the radiation field
obtained from the conformal treatment by restriction to $\scri^+$ and
introduction of a (special) Bondi parameter exactly reproduces the
radiative information that we are interested in.

However, in most cases of interest there is no preferred Bondi
parameter and then we have to regard all parameters as equivalent. To
see what changes when a different parameter is used, we recall that
any Bondi parameter is of the form $\bar u=Gu+H$. Since the shift of
origin is of no consequence we choose $H=0$. Then $u=\bar u/G$ and
\begin{equation}
  i^*F=- \frac1G\,\ddot {p}(\bar u/G) \, d\bar u\wedge dy ,
\end{equation}
so that the radiation field expressed in terms of $\bar u$ is
$\frac1G\,\ddot {p}(\bar u/G)$.

In order to interpret this behaviour we have to keep in mind that the
electromagnetic field has been given with respect to one specific
Minkowski coordinate system. But all such systems are
equivalent. Suppose we choose a new Minkowski coordinate system which
is boosted with velocity $w$ with respect to the old one along the
direction towards the detector. Then one finds the transformation of
the retarded time to be $u\mapsto \bar u=\alpha u$ with
$\alpha=\sqrt{(1+w)/(1-w)}$. The radiative components of the covariant
fields transform according to
\begin{equation}
 {(E_{\mbox{rad}},B_{\mbox{rad}}) } \mapsto {(\bar E_{\mbox{rad}}, \bar
   B_{\mbox{rad}}) } = \frac1\alpha {(E_{\mbox{rad}},B_{\mbox{rad}}) }.
\end{equation}
For $E_{\mbox{rad}}$ we then have
\begin{equation}
  E_{\mbox{rad}} \mapsto \frac1\alpha E_{\mbox{rad}} = -\frac1\alpha \ddot
  p(u)\,dy =  -\frac1\alpha \ddot
  p(\bar u/\alpha)\,dy
\end{equation}
and similarly for $B_{\mbox{rad}}$. Identifying $G$ and $\alpha$ we
have the same transformation law. This means that the freedom in the
Bondi parameters reflects the freedom in the choice of Minkowski
coordinates. Mathematically, this is to be expected and physically,
this is a well known phenomenon, namely the Doppler shift in the
incident signal due to the relative motion of the source and the
detector. The ambiguity arises because this motion is not known a
priori but has to be determined by other means (e.g., spectroscopic
methods).

\section{Gravitational radiation}
\label{sec:gravrad}

Let us now see how the above discussion applies to the more
complicated case of gravity. We have already mentioned above that the
concept of the detector at infinity as stated in
section~\ref{sec:detect} is entirely intrinsic to $\scri$. It can be
applied whenever there is a well-defined notion of null-infinity
available. Therefore, we only need to see what corresponds to the
radiative information which, in our simple example of the Maxwell
case, was given by the changing dipole moment. For the following
discussion we refer to the work of Janis and
Newman~\cite{NewmanJanis-1965} on the structure of the sources of the
gravitational field. This paper contains a unified treatment of
Maxwell theory and linearized gravity which is very useful for our
purposes. Incidentally, these authors treat the radiation of the
general multipole moments from which it becomes obvious that our
discussion which was restricted to the dipole case generalizes in a
straightforward way without any change in the essential statements.

We put the electromagnetic dipole field in a form which is very much
like the one for the gravitational field. We introduce the Bondi frame
of (complex) null vector fields
\begin{equation}
  \mathbf{l}=2\del_v,\qquad \mathbf{n}=\del_u,\qquad 
  \mathbf{m}=\frac{1}{\sqrt{2}r}\left(\del_\theta 
    + \frac{i}{\sin\theta} \del_\phi \right).
\end{equation}
Then we can encode the electric and magnetic fields in three complex
scalar functions $\phi_0$, $\phi_1$, $\phi_2$ as follows:
\begin{align}
  \phi_0 &= F(\mathbf{l},\mathbf{m}) = \frac1{\sqrt2}\frac{p}{r^3}\,
  \sin\theta ,\\
  \phi_1 &= \frac12\left(F(\mathbf{l},\mathbf{n}) +
    F(\mathbf{m},\mathbf{\bar m}) \right) = - \left(
    \frac{p}{r^3} + \frac{\dot p}{r^2} \right)\cos\theta ,\\
  \phi_2 &= F(\mathbf{\bar m},\mathbf{n}) = -\frac1{\sqrt2}
  \left(\frac{\ddot p}{r} + \frac{\dot 
      p}{r^2} + \frac{p}{2r^3} \right) \sin\theta
\end{align}
A simple definition of $a_1=p/\sqrt2$ achieves complete agreement with
the dipole field given in~\cite{NewmanJanis-1965}. These authors define
 the coefficient $\phi_2^0(u,\theta,\phi)$ of $1/r$ in the scalar
$\phi_2$ as the ``news function''. Physically, this function
is the information sent by a ``broadcasting'' station. Mathematically,
it is one piece of the free data for the solution of the
characteristic initial value problem for the Maxwell equations on
Minkowski space. It is that piece of data which can be considered as
being given on $\scri^+$, or, physically, that component of the field
which can be extracted from $\scri^+$ by an infinite detector. We will 
also call it the ``null-datum'' on $\scri^+$.

The similarity between the Maxwell equations and the equations for
linearized gravity suggests that also in the latter case there is a
``news function''. Indeed, the function in question is $\psi_4^0$, the
component of the linearized Weyl spinor which is entirely intrinsic to
$\scri^+$. It plays the same role for the solution of the
characteristic initial value problem as does the $\phi_2^0$ in the
Maxwell case. Thus, in the linearized gravity case things are very
much in analogy to the Maxwell case. In order to extract the radiative
information one determines the null-datum $\psi_4^0$ along all the
generators of $\scri^+$ as a function of a Bondi parameter.

For the full theory we can again employ an analogy. The asymptotic
structure of the non-linear gravitational field is very similar to
that of the asymptotic linearized field. Again, one can isolate a
function $\Psi_4^0$, the component of the Weyl spinor which is
intrinsic to $\scri^+$. This can also be given freely in the solution
of the asymptotic characteristic initial value problem (see
e.g.,~\cite{NewmanUnti-1962}). Thus, mathematically, it serves the
same purpose as $\phi_2^0$ and $\psi_4^0$. However, its physical
meaning is not so clear cut as before. $\Psi_4^0$ does give
information about the multipole structure of the gravitational
radiation but the implications for the multipole structure of the
source are not as direct as they were before. The field interacts with
itself due to the non-linearity of the theory, thus creating
additional structure not attributable to the source. Examples of this 
are the critical phenomena in the scalar field collapse and the
quasi-normal ringdown of perturbed black holes. Therefore, there is no
direct relationship between the multipole moments of the source on the 
one hand and those of the gravitational radiation on the
other. But irrespective of these issues, the null-datum in the
non-linear theory is $\Psi_4^0$ and it is the one that contains the
radiative information of the gravitational field.  The ``news
function'' for full gravity  has been defined by
Bondi~\cite{BondivanderBurgMetzner-1962}) in  a different way. It is
related to the null-datum $\Psi_4^0$, which is its $u$-derivative. In
contrast to the null-datum which is a local quantity, the news is well 
defined only globally on $\scri^+$  and therefore it is not as
accessible as $\Psi_4^0$.

\section{Discussion and conlusion}
\label{sec:concl}

In all cases we have discussed, there is a null-datum which
contains information about the structure of the field and its
sources. These functions are those which can be registered on
$\scri^+$ by the infinite detectors. The extraction process consists
of evaluating the news on $\scri^+$ as a function of a Bondi parameter
on all generators. Then one obtains the two polarizations of the
radiation which arrives on $\scri^+$. We have described an idealised
extraction process and we have argued that it is a natural
idealisation of the real physical measurement. Furthermore, it is a
procedure which is entirely intrinsic to $\scri^+$, a property which
makes it very valuable within the framework of the conformal
techniques in numerical relativity. 

Within these methods the extraction process becomes an unambiguous
algorithm which allows the determination of the well defined news
function up to the freedom of the choice of an origin of time for each
detector and the indeterminacy due to the unknown boost between the
source and the detector. The only approximation involved is the basic
assumption that the arena be an asymptotically flat space-time. In
contrast to this, other extraction methods which are based on points
in the finite region of space-time have to make additional assumptions
in order to separate the far-field from the near-field.

We have shown in~\cite{jf-1997-3} that one can accurately recover the
news from $\scri^+$ so that the discussion of the present paper is not 
academic but a practical one. What remains to be tested numerically is
the location of the generators of $\scri^+$ on each time-slice and the
determination of the Bondi parameters on them. This will be discussed
in a later paper.


\end{document}